# Relevance of the purity level in a MetalOrganic Vapour Phase Epitaxy reactor environment for the growth of high quality pyramidal site-controlled Quantum Dots

V. Dimastrodonato, L.O. Mereni, R. J. Young[1] and E. Pelucchi

*Tyndall National Institute, University College Cork, Cork, Ireland*

Corresponding author: e-mail valeria.dimas@tyndall.ie, Phone: +353 21 420 4195, Fax: +353 21 4270271, address: Tyndall National Institute "Lee Maltings", Dyke Parade, Cork, Ireland

**Abstract**

**We report in this work on the spectral purity of pyramidal site-controlled InGaAs/AlGaA Quantum Dots grown by metalorganic vapour phase epitaxy on (111)B oriented GaAs substrates. Extremely sharp emission peaks were found, showing linewidths surprisingly narrow (~27μeV) and comparable to those which can be obtained by Molecular Beam Epitaxy in an ultra-high vacuum environment. A careful reactor handling is regarded as a crucial step toward the fabrication of high optical quality systems.**

**Keywords:** A3.Organometalic vapour phase epitaxy, B1.Arsenates, B2.Semiconducting III-V materials, A1.Nanostructures

**PACS:** 68.65.Hb , 71.35.Cc , 78.67.Hc

---

[1] Department of Physics, Lancaster University, Lancaster, LA1 4WY, UK

## 1. Introduction

Semiconductors Quantum Dots (QDs) are often addressed to in the scientific community as artificial atoms, due to the three-dimensional confinement of the charge carriers which leads to discretized energy levels. This inherent property makes QDs ideal candidates as single [1] and entangled [2] photon emitters and intriguing topic of study, especially for their potential applications in quantum information and computing field [3]. An ideal QD system though is required to satisfy some key prerequisites for practical applications. Uniformity and reproducibility of both the growth process and electro-optical characteristics are highly desirable, for example to integrate differently engineered structures on the same chip. A fine control of the nucleation site and density of the structures on the substrate is also required: with particular regard to quantum electrodynamics applications in fact the radiative behaviour of QDs relies not only on the spectral but also on the spatial alignment to the modes present in the cavity [4]. Moreover an easy tuning of the electro-optical properties, possibly achievable through a simple control of some growth parameters, would enable the successful employment of these systems across many different fields.

The development of site control techniques in the last decade partially allowed fulfilling the abovementioned requirements, but a main drawback limits the large-scale employment of QDs controlled in situ as potential building blocks of quantum devices: the pre-growth chemical processing of the substrate , necessary to obtain a uniform template optimized according to a specific application, inevitably introduces sources of contamination, leading as final result to a poorer purity of the spectrum emitted from the samples in comparison with Stranski-Krastanov (SK) grown counterparts. Besides in MetalOrganic Vapour Phase Epitaxy (MOVPE) processes the achievement of high quality represents a critical issue when compared to Molecular Beam Epitaxy (MBE): both the purity levels of the group III and V sources and the reactor pressure, far from ultra-high vacuum conditions, are the main factors hindering the fabrication of structures with ultra low background doping incorporation. An effort is then essential to control the impurities inside the reactor and eventually propose efficient solutions to reduce them.

Pyramidal QDs have revealed in the last years to be, in comparison to other site-controlled techniques [5], an excellent approach toward the compromise spectral purity/uniformity [6]. Their growth mechanism can in fact deliver an enhanced spectral purity due to the localization of the dot which is buried in between relatively thick epitaxial layers, far from the patterned interface. Nevertheless the incorporation of unintentional impurities prevented, in the past, the attainment of very narrow exciton peak linewidths, limiting the Full Width at Half Maximum (FWHM) values to ~100μeV.

We have recently demonstrated [7] that InGaAs QD in GaAs barriers can show dramatically improved optical qualities, being the best FWHM measured from a 0.5 nm thick system with 25% of In equal to a record ~18μeV. It has been widely observed that in MOVPE processes GaAs layers present a lower probability to incorporate impurities, especially $O_2$ and C [8]. On the other hand it is more problematic to control and therefore reduce the contamination in AlGaAs layers [9]. Although it can always be assumed that an optimization of the growth conditions can deliver better optical quality, we believe that a more general approach is required to assure an improvement of the epitaxial purity, focused on source materials and purification techniques to achieve reduced unintentional doping background in an MOVPE reactor. Bearing this in mind we have recently been able to grow GaAs quantum wells in AlGaAs barriers with record properties [10], together with the already cited pyramidal site controlled QDs, within the InGaAs in GaAs barriers system, exhibiting excellent optical quality [7]. We show here that these performances can be extended to other pyramidal QD systems, and in particular we present preliminary results obtained with the more challenging InGaAs in AlGaAs barriers QDs, showing PhotoLuminescence (PL) peaks with linewidths remarkably narrower than their previously grown counterparts [11,12].

We briefly discuss in the following section the key parameters which need to be controlled in order to carry out this approach and we describe in detail the main structural and optical properties of pyramidal site-controlled QDs grown under the adopted procedure.

## 2. Experimental procedure

All our samples are grown by MOVPE in a commercial Aixtron 200 1x2" horizontal reactor with Nitrogen as carrier gas and using standard precursors trimethyl-gallium/aluminium/indium, for group III sources, and arsine, as group V material. An extremely low contamination level environment characterizes our reactor: a double purifier system, consisting of two different hydride purifiers put in series, allowed us to reach a purity level, regarding arsine, <<1ppb ($H_2O$, $O_2$) [10]. Low pressure (20mbar) conditions are employed in order to further reduce the presence of impurities and all growths are performed with our reactor inner walls baked and coated with material from previous runs. We carefully monitor our contamination background by routine fabrication of our high quality GaAs/AlGaAs Quantum Well (QW) structures, which emit, under optimized conditions, neutral excitons with a linewidth as narrow as ~0.41meV [10], value very close to the record obtained by MBE. This is the fundamental requisite enabling the excellent optical quality of our pyramidal QD systems.

Pyramidal site-controlled QDs are grown on GaAs substrates oriented along the direction (111)B, patterned prior to growth through standard photolithography, in order to obtain a uniform template of inverted pyramidal recesses, with a pitch of 7.5μm, acting as nucleation site during growth. A basic pyramidal QD system consists in the epitaxial deposition of: a GaAs buffer, trapping the organic contaminations deriving from the previous chemical steps of the substrate patterning; an etch-stop $Al_{0.8}Ga_{0.2}As$ layer, which allows a selective post-growth chemical etching of the substrate in a process known as back-etching [13,14]; a double barrier structure, composed of an outer layer of $Al_{0.55}Ga_{0.45}As$, which helps confining the carriers in the dot, and an inner layer of (Al)GaAs, providing the confinement. The dot layer material is InGaAs, whose In concentration and thickness can be finely modulated according to the required emission wavelength. Growth temperatures are in the range 730ºC-800ºC (thermocouple reading, the higher temperature is employed in order to reduce the incorporation of unintentional doping, especially during the deposition of AlGaAs layers) and V/III ratios vary from 600 up to 800.

These growth conditions are optimized to assure a reliable formation of the dot in the inverted pyramid. During the layer deposition in fact the original tip of the structure, which chemical selective etching delivers very sharp, undergoes an evolution toward a wider base as a consequence of two effects: decomposition rate anisotropy and capillarity [15]. In particular due to the different chemical bonds exposed on the lateral (111)A facets and the (111)B oriented bottom, a higher growth rate will be obtained on the walls, resulting in a very sharp tip. The shape of the recess on the other hand leads to a broadening of the tip of the pyramid due to a higher growth rate, resulting from the diffusion of the adatoms from the lateral facets toward the bottom. Under critical growth conditions equilibrium between these two effects can be reached and the formation of a self-limited profile occurs along the vertical central axis of the structure. The dot can form therefore only when the base of this profile is narrow enough to assure a lateral confinement of the dot layer deposited between the barriers.

Fig.1 shows a representative cross-section Atomic Force Microscopy (AFM) scan of a superlattice structure where the formation of the self-limited profile can be clearly observed along the axis of the inverted pyramid. The sample under investigation consists of a GaAs buffer, four periods of 21nm thick $Al_{0.3}Ga_{0.7}As$ and 6 nm thick GaAs layers separated by thicker (15nm) GaAs markers, labelled in figure, and a GaAs cap. The base of the pyramid, outlined with the dashed line, is initially broader, due to a pronounced capillarity effect in the GaAs layer (as consequence of a large diffusion of Ga adatoms toward the bottom), but then evolves toward a narrower profile after a critical thickness, marked with a transversal line, where decomposition rate anisotropy and capillarity reach the equilibrium. The same figure offers an example of vertically stacked nanostructures: with this approach several QDs can be in fact aligned along the vertical axis. The extremely versatile growth mechanism (compared to SK growth) allows a fine modulation of the thickness/concentration of dots and barriers, opening a broad range of opportunities for the fabrication of differently engineered coupled QD systems.

## 3. Results and discussion

During the MOVPE growth process the original triangular top shape of the structure changes into a hexagon [14,16], due to the evolution of the as-etched (111)A lateral planes into vicinal facets. Fig.2 shows a typical top view scan, performed by AFM, of an as-grown InGaAs/GaAs pyramidal QD system with pure GaAs barriers (we caution the reader that the top shape of the structure presents always similar profiles both in the case of InGaAs/AlGaAs and InGaAs/GaAs systems, since it is the result of the evolution of the final AlGaAs cladding layer lateral walls and function of the thickness of the layers only). The outer triangular etched profile is visible and highlighted with a dashed line. The inner hexagonal shape well points out the evolution of each (111)A lateral walls into two vicinal (111)A facets. The bottom of the pyramid, although sharp, appears flat due to the AFM scanning tip which does not reach the base at the bottom of the structure. In the inset a High Resolution Scanning Electron Microscopy (HRSEM) top view of an *as-etched* single pyramid (prior to growth) is shown to compare the original shape with the *as-grown* profile.

After growth all samples undergo further chemical processes known respectively as surface- and back-etching [13]: the former allows us to remove irregular facets which tend to form during growth between the planar top (111)B substrate and the lateral (111)A walls, the emission of which might hide that from the dot; the latter, as already anticipated, is a selective chemical removal of the substrate, sometimes necessary to improve the collection efficiency of the PL signal emitted from the specimen[14].

The samples are then mounted on a cold finger of a closed cycle Helium cryostat and their emission spectra are measured at low temperature, ~10K, with a micro-PL set-up, by exciting a single pyramidal dot with a laser emitting at 658nm and acquiring the emitted signal with a CCD camera.

A 0.5nm thick $In_{0.1}Ga_{0.9}As/Al_{0.3}Ga_{0.7}As$ QD system has been characterized in back etching geometry with our micro-PL set-up to investigate the optical properties: most of the pyramidal dots emit spectra with exciton peak linewidths in the range 40μeV-70μeV; nevertheless some of them exhibit extremely narrow PL peaks.
In Fig.3 we show one of the sharpest excitonic peaks emitted from the sample under investigation: the correspondent FWHM value of ~27μeV, extrapolated from the Lorentzian fitting of the PL peak, is indicative of high crystal and optical quality of the system grown in our reactor. This striking result represents an important improvement achieved in terms of spectral purity in InGaAs/AlGaAs systems: such dots in fact have been extensively grown in the past years, but the best PL peaks exhibited linewidths broader than 80μeV[12]. It is worth stressing here that the spectral purity reached is result mainly of the reactor handling, as already discussed, but a careful attention has to be paid also to the substrate patterning processing, in order to reduce the organic contamination deriving from the chemical steps.

Further optics measurements were conducted to characterize our sample in terms of Fine Structure Splitting (FSS) (polarization dependent spectra were acquired by using a linear polarizer at fixed polarization direction at the entrance of the monochromator and a half-waveplate which was rotated to resolve the linear polarization of the incident beam [17]). FSS is a key feature for a reliable generation of entangled photons. It has been demonstrated that entangled photons can be produced by the biexciton-exciton cascade of a single semiconductor QD: the process releases two photons which, in ideal conditions, share an inseparable wave function [18,2]. Most of the attention so far has been concentrated on self-assembled SK QDs, but their intrinsic in-plane anisotropy and built-in strain results in a break-up of the two fold degeneracy of the exciton state, FSS, which must be very small to avoid entanglement-destruction upon time-averaging [19]. Pyramidal site-controlled QDs grown on GaAs (111)B oriented substrates would ideally overcome such difficulties: the substrate orientation (111) and the patterning process assure a threefold $C_{3v}$ symmetry, ideally sufficient to guarantee a nearly zero QD anisotropic exchange splitting [20]. Moreover, unlike SK systems, pyramidal QDs can form in absence of any built-in strain, resulting in a further reduction of the number of sources which may induce a large exchange splitting.

Nevertheless, preliminary measurements of our InGaAs in AlGaAs QDs revealed values of FSS in the range 30μeV-60μeV. Further improvements in the growth parameters and the lithography protocols are though needed to reduce the splitting and more detailed analyses are required in order to understand the reason of such relatively large values.

## 4. Conclusions

We have discussed in this manuscript the importance of achieving high optical quality of MOVPE grown structures, with particular regard to pyramidal site-controlled QDs. Record spectral purity was obtained with InGaAs/AlGaAs QD systems, exhibiting sharp exciton PL peaks with FWHM values remarkably lower than the previous analogue dots.

## 5. Acknowledgments

This research was enabled by the Irish Higher Education Authority Program for Research in Third Level Institutions (2007-2011) via the INSPIRE programme, and by Science Foundation Ireland under grants 05/IN.1/I25. We are grateful to K. Thomas for his support with the MOVPE system.

**Figure captions**

Fig.1: Cross-section of a superlattice AlGaAs/GaAs pyramidal structure scanned by AFM. Thicker GaAs markers are labelled and the evolution of the wide base toward a narrower self-limited profile can be observed along the vertical axis of the pyramid. Note that, although the nominal thickness of the layers is kept constant during growth, a faster dynamics of the decomposition/incorporation processes occurs as we fill the pyramid, leading to an actual higher growth rate of the upper layers.

Fig.2: AFM scan in top view of a typical InGaAs/GaAs pyramidal QD. The as-etched triangular top shape is outlined with a dashed line and the as-grown hexagonal profile, caused by the evolution of the original (111)A walls toward the vicinal facets, is clearly visible. In the inset a HRSEM image of the etched pyramidal recess is shown to better compare the triangular shape with the modified profile obtained after growth.

Fig. 3: Low temperature emission spectrum of a single 0.5nm thick InGaAs/AlGaAs pyramidal site-controlled QD. The value of the FWHM of the peak has been extrapolated from a Lorentzian fitting of the measured emission and marks a remarkable improvement in terms of spectral purity with regard to AlGaAs based QD systems.

V. Dimastrodonato et al.

**Figure 1**

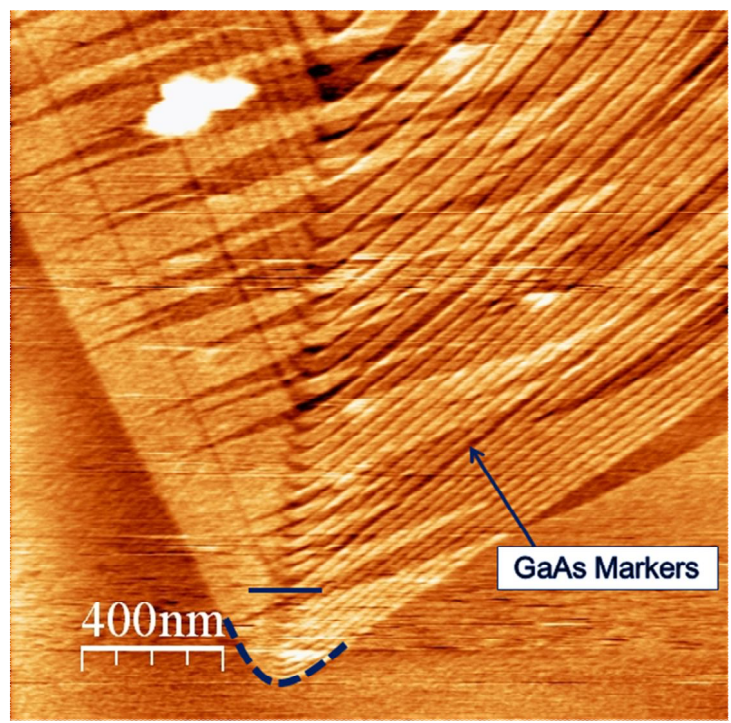

**Figure 2**

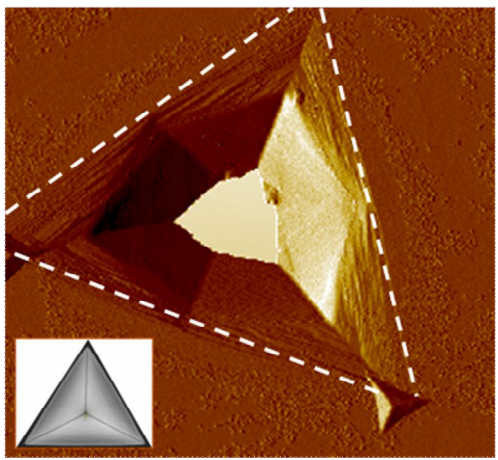

**Figure 3**

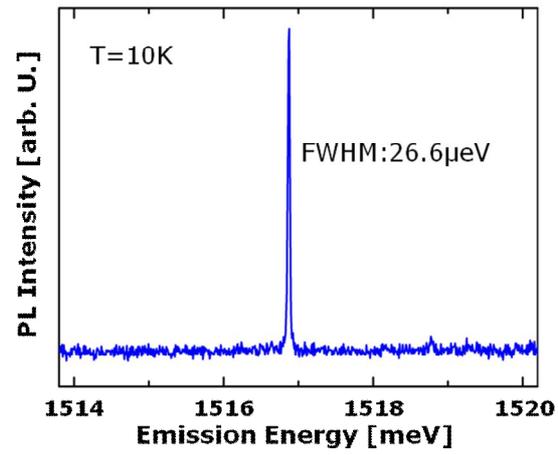